\documentclass[11pt,journal,draftcls,onecolumn,peerreviewca]{IEEEtran}

\usepackage{url}

\usepackage{graphicx}

\usepackage{cite}
\usepackage{amsmath,amssymb,amsfonts}
\usepackage{algorithm}
\usepackage{algorithmic}
\usepackage{graphicx}
\usepackage{textcomp}
\usepackage{xcolor}
\usepackage{amsfonts}
\usepackage{amssymb}
\usepackage{bm}
\usepackage{color, soul}
\usepackage{stfloats}
\usepackage[numbers,sort&compress]{natbib}
\usepackage[amsmath,thmmarks]{ntheorem}
\usepackage{theorem}
\usepackage{subfigure}
\usepackage{ntheorem}
\usepackage{url}
\usepackage{hyperref}
\usepackage{float}
\usepackage{mathrsfs}
\usepackage{algorithm}
\usepackage{algorithmic}
\usepackage{mathrsfs}

\begin{document}

\title{Multidimensional Variational Line Spectra Estimation}
%
%\author{Qi Zhang, Jiang Zhu, Ning Zhang and Zhiwei Xu}
%\thanks{Q. ZHang, J. Zhu and Z. Xu are with the Ocean College, Zhejiang University, Zhoushan, 316021, China. N. Zhang is with Nanjing Marine Radar Institute, Jiangning District, Nanjing, 211153, China.}
%
%\markboth{Journal of \LaTeX\ Class Files, Vol. 14, No. 8, August 2015}
%\maketitle

\author{Qi Zhang, Jiang Zhu, \IEEEmembership{Member, IEEE}, Ning Zhang and Zhiwei Xu, \IEEEmembership{Senior Member, IEEE}
%\thanks{This work was supported in part by the NSFC under Grant 61901415}
\thanks{Q. Zhang, J. Zhu and Z. Xu are with the Ocean College, Zhejiang University, Zhoushan, 316021, China (e-mail: \{zhangqi13, jiangzhu16, xuzw\}@zju.edu.cn).}
\thanks{N. Zhang is with Nanjing Marine Radar Institute, Jiangning District, Nanjing, 211153, China (e-mail: zhangn\_ee@163.com).}}

\markboth{Journal of \LaTeX\ Class Files, Vol. 14, No. 8, August 2015}
{Shell \MakeLowercase{\textit{et al.}}: Bare Demo of IEEEtran.cls for IEEE Journals}
\maketitle

\begin{abstract}
The fundamental multidimensional line spectral estimation problem is addressed utilizing the Bayesian methods. Motivated by the recently proposed variational line spectral estimation (VALSE) algorithm, multidimensional VALSE (MDVALSE) is developed. MDVALSE inherits the advantages of VALSE such as automatically estimating the model order, noise variance and providing uncertain degrees of frequency estimates. Compared to VALSE, the multidimensional frequencies of a single component is treated as a whole, and the probability density function is projected as independent univariate von Mises distribution to perform tractable inference. Besides, for the initialization, efficient fast Fourier transform (FFT) is adopted to significantly reduce the computation complexity of MDVALSE. Numerical results demonstrate the effectiveness of the MDVALSE, compared to state-of-art methods.
\end{abstract}

%\begin{IEEEkeywords}
{\bf keywords}: Variational Bayesian inference, line spectral estimation, von Mises distribution, multidimensional frequency estimation
%\end{IEEEkeywords}

%\IEEEpeerreviewmaketitle

\section{Introduction}

The problem of multidimensional (MD) frequency estimation arises in many applications. For the $2$-D frequency estimation in wireless communication systems, the two dimensions correspond to the delay and angle of each path. Physical arguments and a growing body of experimental evidence suggest that utilizing the structures benefits and improves the channel reconstruction performance \cite{Jinshi}. Another application is for the $3$-D frequency estimation in radar systems, one can interpret one dimension as time delays, one dimension as Doppler shifts, and the other as spatial frequencies. Since each pair of time delay, Doppler shift and spatial frequency specifies the localization, speed and direction of arrival (DOA) of a scatter, respectively, estimating these parameters is of great importance for target localization and tracking \cite{Jin}. Traditional methods including $N$-D Discrete Fourier Transform (DFT), subspace based approaches such as $2$D MUSIC \cite{2DMUSIC}, $2$D ESPRIT \cite{2DESPRIT}, Matrix Enhancement Matrix Pencil (MEMP) \cite{MEMP} and the multidimensional folding (MDF) techniques \cite{MDF1, MDF2}.

%For $N$-D frequency estimation, an efficiently implementable $N$-D Discrete Fourier Transform (DFT) is usually adopted. However, DFT suffers from off-grid effects \cite{Chi1} and inter-target interference, which results in poor estimation performance and low resolution \cite{Madhow}. As for the subspace based approaches, especially for $2$-D frequency estimation, $2$D MUSIC \cite{2DMUSIC}, $2$D ESPRIT \cite{2DESPRIT}, Matrix Enhancement Matrix Pencil (MEMP) \cite{MEMP} and the multidimensional folding (MDF) techniques \cite{MDF1, MDF2} are proposed. However, the performance of the subspace based approaches rely on the knowledge of the model order (the number of sinusoids) and are sensitive to model mismatch and noise \cite{ChiC2D}.

Recently, compressed sensing based approaches have been proposed for $N$-D frequency estimation, especially for $2$-D frequency estimation. Provided that the frequencies lie on the DFT grid, the signal can be recovered via efficient $l_1$ minimization \cite{Candes}. However, due to spectral leakage of the off-grid frequencies, the signal is not exactly sparse and conventional compressed sensing (CS) algorithms may result in significant performance degradation \cite{Chi1}. Consequently, grid refinement and continuous CS approaches have been developed.  In \cite{NOMP2D, Madhow}, the Newtonalized orthogonal matching pursuit (NOMP) is proposed for $2$-D frequency estimation, which first detects the grid and then iteratively refines the frequencies. Besides, the model order is estimated by the constant false alarm (CFAR) criterion with the knowledge of the noise variance. The computation complexity of NOMP is low, and the estimation performance is good. For the latter, \cite{RsCS} proposes a nuclear norm minimization over multi-fold Hankel matrices approach, \cite{yangzaiMD} develops a reweighted trace minimization (RWTM) plus matrix pencil and (auto-)pairing (MaPP) method, where RWTM fully exploits the multilevel Toeplitz (MLT) structure and utilizes the sparsity to obtain the covariance estimates, and MaPP calculates the Vandermonde decomposition. As \cite{RsCS, yangzaiMD} involve solving a semidefinite programming (SDP), their computation complexities are very high.

A more recent approach is the variational line spectral estimation (VALSE) algorithm, which uses a complete Bayesian treatment by imposing the frequencies as random variables \cite{Badiu}. Such method naturally allows for representing and operating with the uncertainty of the frequency estimates, in addition to only that of the weights. Numerically, accounting for the frequency uncertainty benefits the line spectral estimation performance. VALSE automatically estimates the model order, noise variance, and provides the uncertain degrees of the frequency estimates. Besides, the computation is low, compared to the nuclear norm minimization based approaches. It was also extended to the multi-snapshot case in array signal processing \cite{MVALSE} and quantized setting in high-speed sampling \cite{VALSEQ}. In contrast, this work develops MDVALSE to deal with the multidimensional (MD) frequency estimation problem.

The main contribution of this work is as follows: Firstly, MDVALSE is rigorously developed, which is nontrival from the following two aspects: one is that the multifrequencies are treated and updated as a whole, another aspect is that even the probability density function (PDF) of the multifrequencies can be approximated as a multivariate von Mises distribution, computing expectations of $\prod_i{\rm e}^{{\rm j}m_i\phi_i}$ is still hard. We novelly project the PDF of the multifrequencies as independent vonMises distribution, which allows computation of $\prod_i{\rm e}^{{\rm j}m_i\phi_i}$. Secondly, the $N$-D fast Fourier transform (FFT) approach is adopted to reduce the computation complexity significantly. Finally, numerical experiments are conducted to demonstrate the excellent performance of MDVALSE.

\section{Problem Setup}

Consider a measurement model for $D$ dimensional line spectra estimation problem expressed as
\begin{align}\label{signal-model}
{\mathbf Y}={\mathbf X}+{\mathbf N},
\end{align}
where ${\mathbf X}$ is the noiseless line spectral composed of $K$ complex sinusoids
\begin{align}
{\mathbf X}=\sum\limits_{k=1}^K{w}_{k}{\mathbf A}({\boldsymbol\theta}_{k}),
\end{align}
$w_k$ denotes the complex amplitude of the $k$th frequency, ${\mathbf A}({\boldsymbol\theta}_k)\in{\mathbb C}^{M_1\times\cdots\times M_D}$ and the $(m_1,\cdots,m_D)$ th element is $\prod_{d=1}^D{\rm e}^{{\rm j}(m_d-1){\theta}_{k,d}}$, ${\theta}_{k,d}\in[-\pi,\pi]$ is the $d$th element of ${\boldsymbol\theta}_k$ and ${\boldsymbol\theta} = [{\boldsymbol\theta}_1,\cdots,{\boldsymbol\theta}_K]$, ${\mathbf N}$ is the additive white Gaussian noise (AWGN) and ${\mathbf N}_{\mathcal M}\sim {\mathcal {CN}}({\mathbf N}_{\mathcal M}; 0,\nu)$, where
\begin{align}
    {\mathcal M} = (m_1,\cdots,m_D).
\end{align}
For simplicity, we define
\begin{align}
a(\mathcal M,{\boldsymbol \theta}_k) = \prod_{d=1}^D{\rm e}^{{\rm j}(m_d-1){\theta}_{k,d}}.
\end{align}

Since the number of spectral $K$ is unknown, an overcomplete model is adopted, where the number of spectral is supposed to be $N$ and $N>K$ \cite{Badiu}. Consequently, the binary hidden variables ${\mathbf s}=[s_1,...,s_N]^{\rm T}$ are introduced to characterize whether the $i$th component is active or not. The probability mass function (PMF) is $p({\mathbf s};\rho)=\prod_{k=1}^Np(s_k;\rho)$, where $s_k\in\{0,1\}$ and
\begin{align}\label{pmfs}
p(s_k;\rho) = \rho^{s_k}(1-\rho)^{(1-s_k)}.
\end{align}
Besides, $p({\mathbf w}|{\mathbf s};\tau)=\prod_{k=1}^Np({w}_{k}|s_k;\tau)$, where $p({w}_{k}|s_k;\tau)$ follows a Bernoulli-Gaussian distribution
\begin{align}\label{pdfws}
p({w}_{k}|s_k;\tau) = (1 - s_k){\delta}({w}_{k}) + s_k{\mathcal {CN}}({w}_{k};0,\tau),
\end{align}
where ${\delta}(\cdot)$ is the Dirac delta function. From (\ref{pmfs}) and (\ref{pdfws}), it can be seen that $\rho$ controls the probability of the $i$th component being active.
For measurement model
(\ref{signal-model}), the likelihood $p({\mathbf Y}|{\mathbf X};\nu)$ is
\begin{align}
p({\mathbf Y}|{\mathbf X};\nu)=\prod\limits_{\mathcal M}{\mathcal {CN}}(Y_{\mathcal M};{X}_{\mathcal M},\nu).
\end{align}
Let ${\boldsymbol \beta} = \{\nu,~\rho,~\tau\}$ and ${\boldsymbol \Phi}=\{{\boldsymbol \theta}, {\mathbf w}, {\mathbf s}\}$ be the model and estimated parameters. In general, one should first find the maximum likelihood (ML) estimates of ${\boldsymbol \beta}$ by marginating the posterior PDF of ${\boldsymbol \Phi}$, then find the maximum a posterior (MAP) or minimum mean squared error (MMSE) estimates of ${\boldsymbol \Phi}$. However, the above approach is computationally intractable and thus an iterative algorithm is developed in the ensuing section.

\section{MDVALSE Algorithm}
The variational approach tries to find a surrogate PDF $q({\boldsymbol \Phi}|{\mathbf Y};{\boldsymbol \beta})$ via minimizing the Kullback-Leibler (KL) divergence between $q({\boldsymbol \Phi}|{\mathbf Y})$ and $p({\boldsymbol \Phi}|{\mathbf Y})$, which is equivalent to maximize
\begin{align}\label{DL-L}
{\mathcal L}(q({\boldsymbol \Phi}|{\mathbf Y});{\boldsymbol \beta}) = {\rm E}_{q({\boldsymbol \Phi}|{\mathbf Y})}\left[\ln{\tfrac{p({\mathbf Y},{\boldsymbol \Phi;\boldsymbol \beta})}{q({\boldsymbol\Phi}|{\mathbf Y})}}\right].
\end{align}
Similar to \cite{Badiu}, $q({\boldsymbol \Phi}|{\mathbf Y};{\boldsymbol \beta})$ is imposed to have the following structures
\begin{align}
q({\boldsymbol \Phi}|{\mathbf Y}) = \prod_{k=1}^Nq(\boldsymbol\theta_k|\mathbf Y)q({\mathbf {w|Y,s}})\delta({\mathbf s}-{\mathbf s}_0).\label{postpdf}
\end{align}

%In Bayesian model, we would like to compute mean and circular mean estimates of $\mathbf w$ and $\boldsymbol\theta$, respectively, based on the posterior PDF
%\begin{align}\label{real-pospdf}
%    p({\boldsymbol \Phi}|{\mathbf Y};{\boldsymbol \beta}) = p({\mathbf Y}|{\mathbf X};\nu)p(\boldsymbol\theta)p(\mathbf w|\mathbf s;\tau)p(\mathbf s),
%\end{align}
%which is intractable for $\mathbf s$ has $2^N$ possible values. Therefore, a surrogate pdf $q({\boldsymbol \Phi}|{\mathbf Y};{\boldsymbol \beta})$ is proposed in VALSE algorithm to approximate (\ref{real-pospdf}) which is  \cite{Badiu}
%\begin{align}
%q({\boldsymbol \Phi}|{\mathbf Y}) = \prod_{k=1}^Nq(\boldsymbol\theta_k|\mathbf Y)q({\mathbf {w|Y,s}})\delta({\mathbf s}-{\mathbf s}_0).\label{postpdf}
%\end{align}
%In addition, VALSE algorithm approximates $q({\boldsymbol \Phi}|{\mathbf Y})$ by minimizing the

Maximizing ${\mathcal L}(q({\boldsymbol \Phi}|{\mathbf Y}))$ with respect to all the factors is also intractable. Similar to the Gauss-Seidel method, $\mathcal L$ is optimized over each factor $q({\boldsymbol\theta}_k|\mathbf Y)$, $k=1,\dots,N$ and $q({\mathbf {w,s|Y}})$ separately with the others being fixed. Let ${\mathbf z}=({\boldsymbol\theta}_1,\dots,{\boldsymbol\theta}_N,({\mathbf w},{\mathbf s}))$ be the set of all latent variables. Maximizing ${\mathcal L}(q({\boldsymbol \Phi}|{\mathbf Y});{\boldsymbol \beta})$ (\ref{DL-L}) with respect to the posterior approximation $q({\mathbf z}_n|{\mathbf Y})$ of each latent variable ${\mathbf z}_n,~n=1,\dots,N+1$ yields
\begin{align}\label{upexpression}
\ln q({\mathbf z}_n|{\mathbf Y})={\rm E}_{q({{\mathbf z}\setminus{\mathbf z}_n}|{\mathbf Y})}[\ln p({\mathbf Y},{\mathbf z})]+{\rm const},
\end{align}
where the expectation is with respect to all the variables ${\mathbf z}$ except ${\mathbf z}_n$ and the constant ensures normalization of the PDF.

Due to the factorization property of (\ref{postpdf}), the frequencies ${\boldsymbol \theta}_k,~k\in\{1,...,N\}$ can be estimated from the marginal distribution $q({\boldsymbol \Phi}|{\mathbf Y})$ as
\begin{subequations}\label{ahat}
\begin{align}
&\widehat{\theta}_{k,d} = {\rm arg}({\rm E}_{q{({\boldsymbol\theta}_k|\mathbf Y})}[{\rm e}^{{\rm j}\theta_{k,d}}]),\label{ahat_a}\\
&\widehat{a}_{\mathcal M,{\boldsymbol\theta}_k} = {\rm E}_{q{({\boldsymbol\theta}_k|\mathbf Y)}}[a(\mathcal M,{\boldsymbol\theta}_k)],~d\in\{1,...,D\},\label{ahat_b}
\end{align}
\end{subequations}
where ${\rm arg}(\cdot)$ returns the angle.

Given that $q({\mathbf s|\mathbf Y}) = \delta({\mathbf s}-{\mathbf s}_0)$, the posterior PDF of $\mathbf w$ is
\begin{align}\label{w_s}
q({\mathbf w}|{\mathbf Y}) = \int q({\mathbf w}|{\mathbf Y},{\mathbf s})\delta({\mathbf s}-{\mathbf s}_0){\rm d}{\mathbf s} = q({\mathbf w}|{\mathbf Y};{\mathbf s}_0).
\end{align}
For the given posterior PDF $q(\mathbf w|{\mathbf Y})$, the mean and covariance of the weights is estimated as
\begin{subequations}\label{w_est}
\begin{align}
&\widehat{\mathbf w} = {\rm E}_{q({\mathbf {w|Y}})}[{\mathbf w}],\\
&\widehat{\mathbf C} = {\rm E}_{q{(\mathbf {w|Y}})}[{\mathbf {w}{\mathbf w}^{\rm H}}] - {\widehat{\mathbf w}}\widehat{\mathbf w}^{\rm H}.
\end{align}
\end{subequations}
Let $\mathcal S$ be the set of indices of the non-zero components of $s$, i.e., $\mathcal S = \{k|1\leq k\leq N,s_k = 1\}$.
%\begin{align}\notag
%\mathcal S = \{k|1\leq k\leq N,s_k = 1\}.
%\end{align}
%Analogously, $\widehat{\mathcal S}$ is defined based on $\widehat{\mathbf s}$. The model order is estimated as the cardinality of $\widehat{\mathcal S}$, i.e.,
%\begin{align}\notag
%\widehat{K} = |\widehat{\mathcal S}|.
%\end{align}
%According to (\ref{signal-model}), the noise-free signal is reconstructed as
%\begin{align}\notag
%\hat{\mathbf X}=\sum_{k\in{\hat{\mathcal S}}}w_{k}{\hat{\mathbf A}_k}.
%\end{align}
In the following, we detail the procedures.

\subsection{Inferring the frequencies}\label{EstFreq}
For each $k = 1,...,N$, we maximize $\mathcal L$ with respect to the factor $q({{\boldsymbol\theta}_k|\mathbf Y})$. For $k\notin{\mathcal S}$, $q(\boldsymbol\theta_k|\mathbf Y)$ remains unchanged.
For $k\in{\mathcal S}$, substituting (\ref{ahat}) and (\ref{w_est}) in (\ref{upexpression}), $\ln q({\boldsymbol\theta}_k|\mathbf Y)$ is obtained as
\begin{small}
\begin{align}\label{pdf-q}
&\ln q({\boldsymbol\theta}_k|\mathbf Y) = \ln p({\boldsymbol\theta}_k)+\sum_{\mathcal M}{\Re}\left\{{\eta}_{k,\mathcal M}a(\mathcal M,{\boldsymbol\theta}_k)\right\} + {\rm const},
\end{align}
\end{small}
where ${\eta}_{k,\mathcal M}$ is
\begin{small}
\begin{align}\label{yita-i}
{\eta}_{k,\mathcal M}=2\nu^{-1}(Y^{*}_{\mathcal M}{{w}_k}-\sum_{i\in\widehat{\mathcal S} \backslash k}{\widehat a}^{*}_{\mathcal M,{{\boldsymbol\theta}_i}}({\widehat C}_{i,k}+{\widehat{w}}_k{\widehat{w}}^{*}_i)),
\end{align}
\end{small}
where $()^*$ denotes the conjugate operation.
While it is hard to obtain the analytical results of (\ref{ahat}) for the PDF (\ref{pdf-q}), $q({\boldsymbol\theta}_k|\mathbf Y)$ is projected as the $D$ independent von Mises distribution. A Newton step is implemented to refine the previous estimates, i.e.,
\begin{subequations}
\begin{align}
&\widehat{\boldsymbol\theta}^{t+1}_ k=\widehat{\boldsymbol\theta}^{t}_k-(\nabla^2f(\widehat{\boldsymbol\theta}^{t}_k))^{-1}\nabla f({\widehat{\boldsymbol\theta}}^{t}_k),\label{theta_post}\\
&\widehat{\boldsymbol\kappa}^{t+1}_k = A^{-1}\left(\exp({2{\rm diag}((\nabla^2f(\widehat{\boldsymbol\theta}^{t+1}_k))^{-1})})\right),\label{kappa_post}
\end{align}
\end{subequations}
where $f({\boldsymbol\theta}_k)$ is the exponential part of $q(\boldsymbol\theta_k|\mathbf Y)$ and the detail of the inverse of $A(\cdot) \triangleq \frac{I_1(\cdot)}{I_0(\cdot)}$ is given in \cite{Direc}, ${\rm diag}(\cdot)$ returns a vector with elements being the diagonal of the matrix. As a result, $q(\boldsymbol\theta_k|\mathbf Y)$ and $\widehat a_{\mathcal M,\boldsymbol\theta_k}$ can be approximated as
\begin{subequations}
\begin{align}
&q(\boldsymbol\theta_k|\mathbf Y) \approx \prod_{d=1}^Df_{\rm VM}(\theta_d;\widehat{\eta}_{k,d}),\label{post_pdf}\\
&\widehat a_{\mathcal M,\boldsymbol\theta_k} \approx \prod_{d=1}^D \frac{I_{{\mathcal M}_d}({\widehat\kappa}_{(k,d)})}{I_0({\widehat\kappa}_{(k,d)})}a_{\mathcal M,{\widehat{\boldsymbol\theta}}_k}.\label{post_mean}
\end{align}
\end{subequations}
In Section \ref{NS}, the accuracy of the approximation (\ref{post_pdf}) is demonstrated numerically.
\subsection{Inferring the weights and support}\label{EstWs}
Next $q({\boldsymbol\theta}_k|{\mathbf Y}),k=1,...,N$ are fixed and $\mathcal L$ is maximized w.r.t. $q({\mathbf w},{\mathbf s}|{\mathbf Y})$.
Define the matrices $\mathbf J$ and $\mathbf H$ as
\begin{subequations}\label{J-H}
\begin{align}
&{J}_{ij}= 	
\begin{cases}
\prod_{d=1}^DM_d,&i=j\\
\sum_{\mathcal M}{\widehat a}^*_{\mathcal M,{\boldsymbol\theta}_i}{\widehat a}_{\mathcal M,{\boldsymbol\theta}_j},&i\neq{j}
\end{cases},\quad i,j\in\{1,2,\cdots,N\},\\
&h_i = \sum_{\mathcal M}Y_{\mathcal M}{\widehat a}^*_{\mathcal M,{\boldsymbol\theta}_i},\quad i\in\{1,\cdots,N\},
\end{align}
\end{subequations}
where ${J}_{ij}$ denotes the $(i,j)$th element of $\mathbf J$. It can be shown that updating $\widehat{\mathbf w}_{{\mathcal S}}$, $\widehat{\mathbf C}_{{\mathcal S}}$ and $\mathbf s$ are the same as \cite{Badiu} and details are omitted here.
%According to (\ref{upexpression}), $q({\mathbf w},{\mathbf s}|{\mathbf Y})$ can be calculated as
%\begin{small}
%\begin{align}
%&\ln q({\mathbf w},{\mathbf s}|{\mathbf Y}) =
%({\mathbf w}_{\mathcal S} - \widehat{\mathbf w}_{{\mathcal S}})^{\rm H}\widehat{\mathbf C}_{{\mathcal S}}^{-1}({\mathbf w}_{\mathcal S} - \widehat{\mathbf w}_{{\mathcal S}}) + {\rm const},\label{Ws_pos}
%\end{align}
%\end{small}where
%\begin{subequations}\label{W-C-1}
%\begin{align}
%&\widehat{\mathbf w}_{{\mathcal S}} = \nu^{-1}\widehat{\mathbf C}_{{\mathcal S}}\mathbf h_{{\mathcal S}},\label{What}\\
%&\widehat{\mathbf C}_{{\mathcal S}} = \left(\frac{{\mathbf J}_{{\mathcal S}}}{\nu}+\frac{{\mathbf I}_{|{\mathcal S}|}}{\tau}\right)^{-1}.\label{defCshat0}
%\end{align}
%\end{subequations}
%It can be shown that updating $\mathbf s$ is the same as \cite{Badiu} and details are omitted here.
\subsection{Estimating the model parameters}
After updating the frequencies and weights, the model parameters $\boldsymbol\beta = \{\nu,~\rho,~\tau\}$ are estimated via maximizing the lower bound ${\mathcal L}(q({\boldsymbol \Phi}|{\mathbf Y});{\boldsymbol \beta})$ for fixed $q({\boldsymbol\Phi}|{\mathbf Y})$. Plugging the postulated PDF (\ref{postpdf}) in (\ref{DL-L}) and set $\frac{\partial\mathcal L}{\partial\nu}=0$, $\frac{\partial\mathcal L}{\partial\rho}=0$, $\frac{\partial\mathcal L}{\partial\tau}=0$, we have
\begin{align}\label{mu-rou-tau-hat}
\widehat{\nu}& = \frac{\left[{\widehat{\mathbf w}}^{\rm H}_{\widehat{\mathcal S}}{\widehat{\mathbf J}}_{\widehat{\mathcal S}}{\widehat{\mathbf w}}_{\widehat{\mathcal S}}+\sum_{\mathcal M}Y_{\mathcal M}^*Y_{\mathcal M}+{\rm tr}(\mathbf J_{\widehat{\mathcal S}}{\widehat {\mathbf C}}_{\widehat{\mathcal S}})-2\Re({\widehat {\mathbf w}}_{\widehat{\mathcal S}}^{\rm H}{\mathbf h}_{\widehat{\mathcal S}})\right]}{\prod_{d=1}^DM_{d}},\notag\\
\widehat{\rho}  &=\frac{||\widehat{\mathbf s}||_0}{N},\quad \quad \widehat{\tau} = \frac{\widehat{\mathbf w}^{\rm H}_{\widehat{\mathcal S}}\widehat{\mathbf w}_{\widehat{\mathcal S}}+{\rm tr}(\widehat{{\mathbf C}}_{\widehat{\mathcal S}})}{||\widehat{\mathbf s}||_0}.
\end{align}

\subsection{Summary of MDVALSE algorithm}
The initialization of MDVALSE algorithm is presented. For the first step of initialization, we randomly extract one dimensional data from $\mathbf Y$ and use the method in \cite{Badiu} to initialize $\widehat{\boldsymbol\beta}$. For the later, we choose to initialize $\{q(\boldsymbol\theta_k|\mathbf Y)\}_{k=1}^N$ in a sequential manner. For the $k$th step, the noncoherent PDF
\begin{align}\label{nonconpdf}
q(\boldsymbol\theta_k|\mathbf Y)\propto \exp\{-|\sum_{\mathcal M}{\mathbf Y}^*_{{\mathcal M},{\rm r}}a(\mathcal M,{\boldsymbol\theta}_k)|^2/(\nu\prod_{d=1}^D M_d)\}
\end{align}
is projected to $D$ independent von Mises distribution, where ${\mathbf Y}_{{\mathcal M},{\rm r}}={\mathbf Y}_{{\mathcal M}}-\sum\limits_{i=1}^{k-1}\widehat{a}_{\mathcal M,{\boldsymbol\theta}_{i}} \hat{w}_{i}$. Here N-D FFT is implemented to obtain the approximated $D$ independent von Mises distribution. Then the MDVALSE can be obtained and the procedures are similar to \cite[Algorithm 3]{Badiu}.

%The MDVALSE is summarized as Algorithm \ref{MDVALSE}.
%In addition, we update the measurements by subtracting the first $k$ signals from $\mathbf Y$ for the next iteration. Thus the MDVALSE is obtained.
%\begin{algorithm}[ht]
%\caption{Outline of MDVALSE algorithm}\label{MDVALSE}
%\textbf{Input:}~~Signal tensor $\mathbf Y$\\
%\textbf{Output:}~~The model order estimate $\widehat{K}$, frequencies estimate $\widehat{\boldsymbol\theta}_{\widehat{\mathcal S}}$, complex weights estimate $\widehat{\mathbf w}_{\widehat{\mathcal S}}$ and reconstructed signal $\widehat{\mathbf X}$
%\begin{algorithmic}[1]
%\STATE Initialize $\widehat{\boldsymbol\nu},\widehat\rho,\widehat\tau$~and~$q(\omega_i|{\mathbf Y}),i\in\{1,\cdots,N\}$
%\STATE \textbf{repeat}
%\STATE ~~~~Calculate $q({\mathbf w},{\mathbf s}|{\mathbf Y})$ and update~$\widehat{\mathbf s},\widehat{\mathbf w}_{\widehat{\mathcal S}}$ (Section \ref{EstWs})
%\STATE ~~~~Update the parameters $\widehat\rho$, $\widehat\tau$ and $\widehat{\nu}$ (\ref{mu-rou-tau-hat})
%\STATE ~~~~Calculate $q({\boldsymbol\theta}|{\mathbf Y})$ and update $\widehat{\boldsymbol\theta}$ (Section \ref{EstFreq})
%\STATE \textbf{until} stopping criterion
%\STATE \textbf{return} $\widehat{K}$, $\widehat{\boldsymbol\theta}_{\widehat{\mathcal S}}$, $\widehat{\mathbf w}_{\widehat{\mathcal S}}$ and $\widehat{\mathbf X}$
%\end{algorithmic}
%\end{algorithm}
% where ${\widehat w}_{i-1}$ is calculated according to (\ref{W-C-1})
\begin{figure}
  \centering
  \subfigure[]{
    \label{2Dcasea}%% label for first subfigure
    \includegraphics[width=90mm]{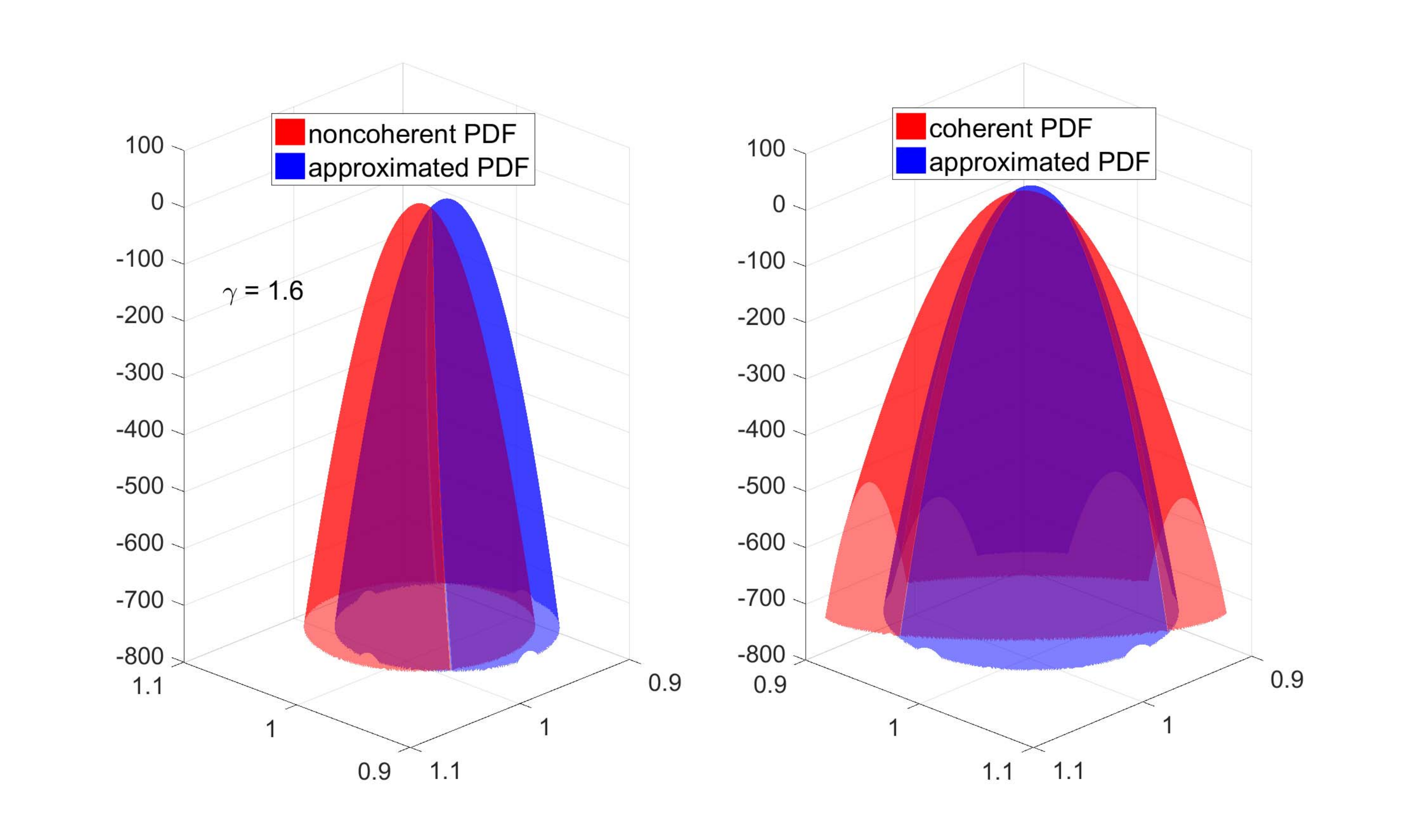}}
  \subfigure[]{
    \label{2Dcaseb}%% label for first subfigure
    \includegraphics[width=90mm]{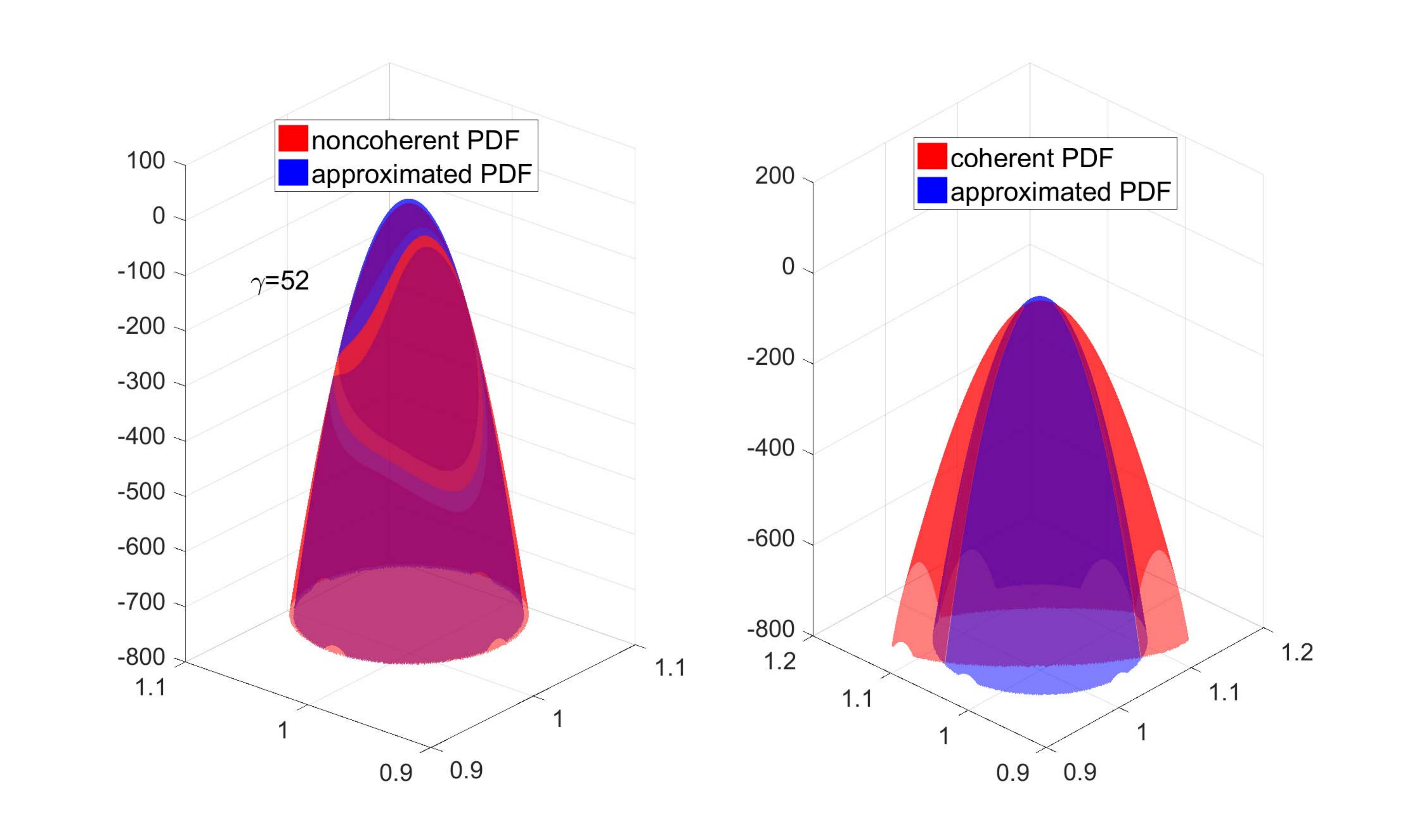}}
 \caption{ The logarithm of the true PDFs (noncohent PDF (\ref{nonconpdf}) and coherent PDF (\ref{pdf-q})) and their approximations for different oversampling factor $\gamma$: (a) $\gamma=1.6$, (b) $\gamma=52$.}
  \label{approx} %% label for entire figure
\end{figure}
\begin{figure}
  \centering
  \subfigure[]{
    \label{2Dcasea}%% label for first subfigure
    \includegraphics[width=2.8in]{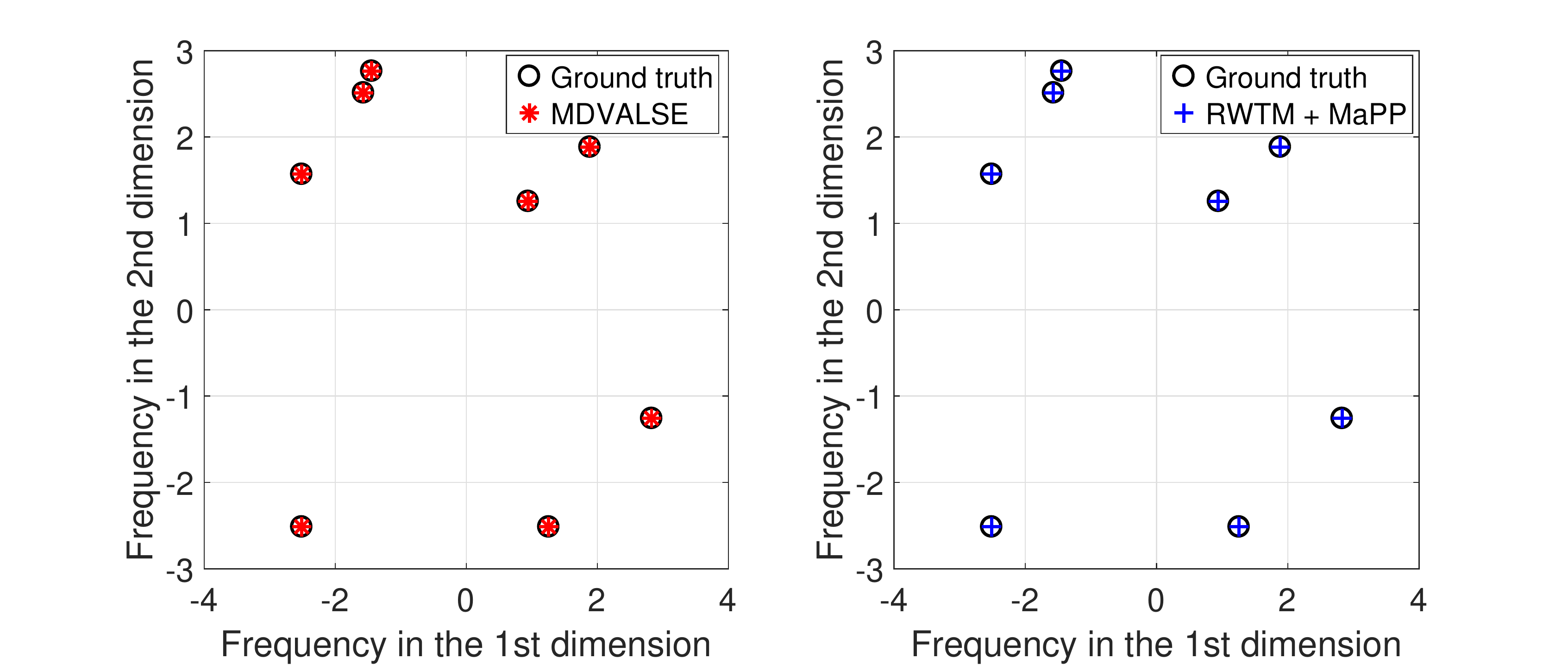}}
  \subfigure[]{
    \label{2Dcaseb}%% label for first subfigure
    \includegraphics[width=2.8in]{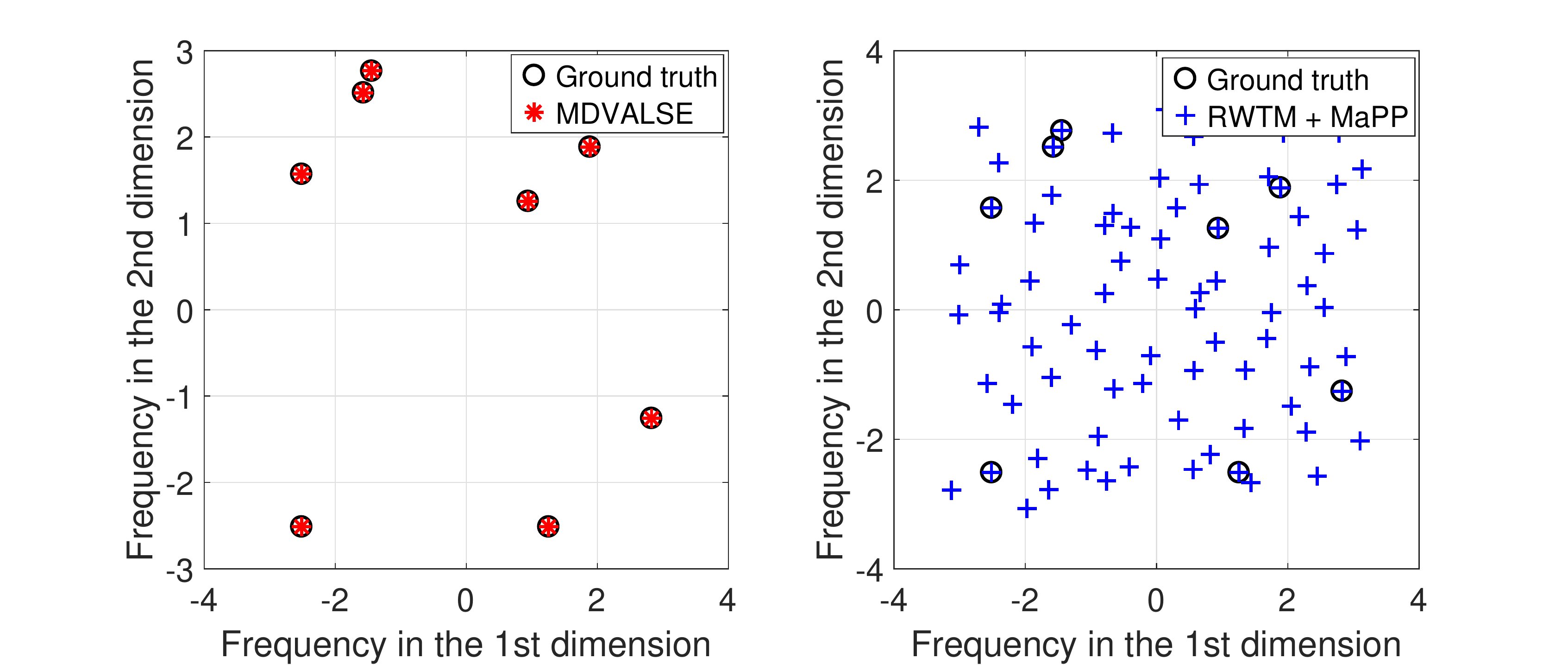}}
 \caption{Recovery performance of RWTM and MDVALSE algorithms for 2D case. (a) Noiseless case and MSE of frequencies is -221 dB and -115 dB for RWTM and MDVALSE, respectively. (b) Noisy case for SNR = 40 dB, and MSE of frequencies is -70 dB for MDVALSE.}
  \label{2Dcase} %% label for entire figure
\end{figure}
\begin{figure*}
  \centering
  \subfigure[]{
    \label{Case2SNRgen1}%% label for first subfigure
    \includegraphics[width=50mm]{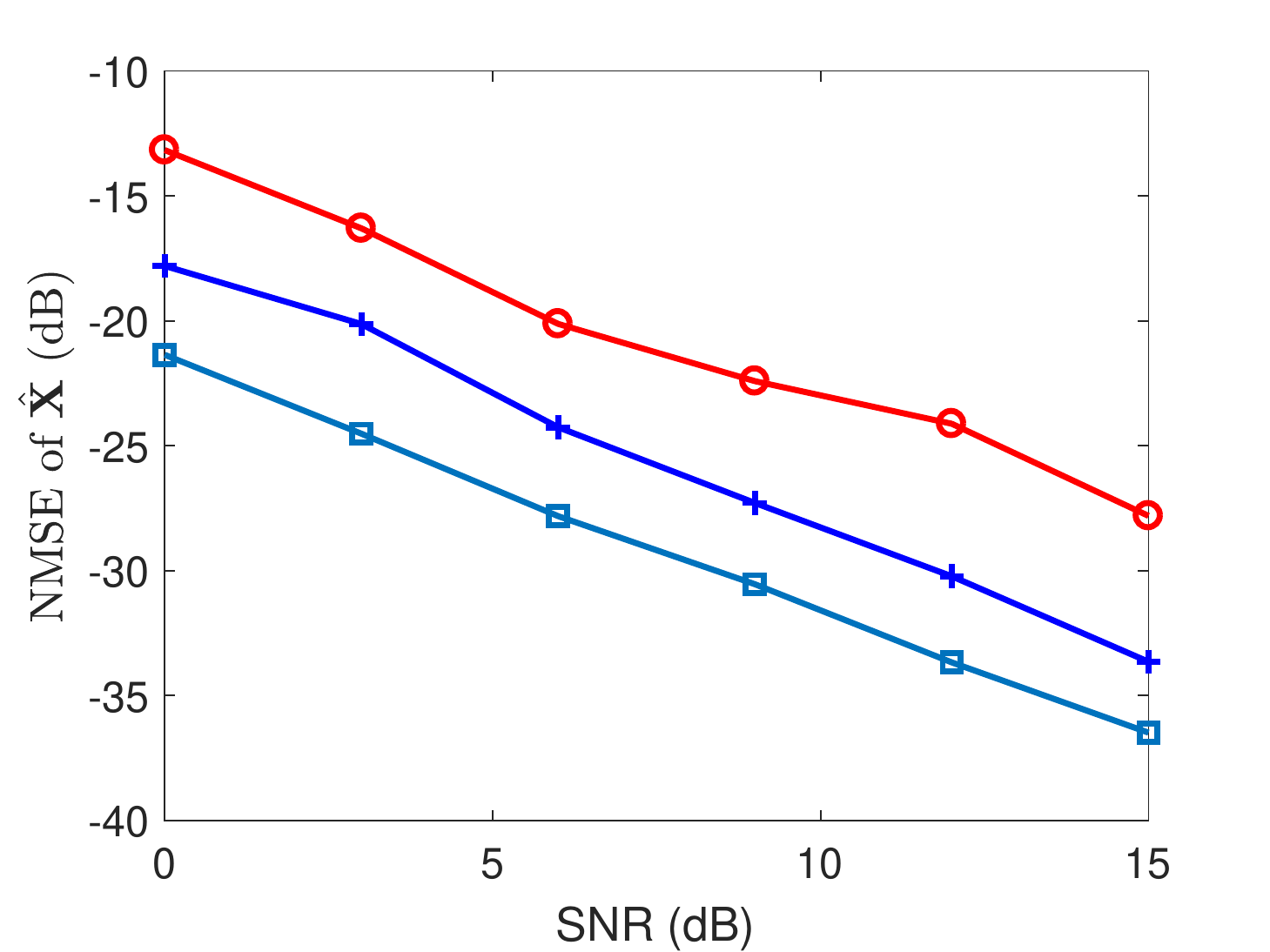}}
  \subfigure[]{
    \label{Case2SNRgen3}%% label for first subfigure
    \includegraphics[width=50mm]{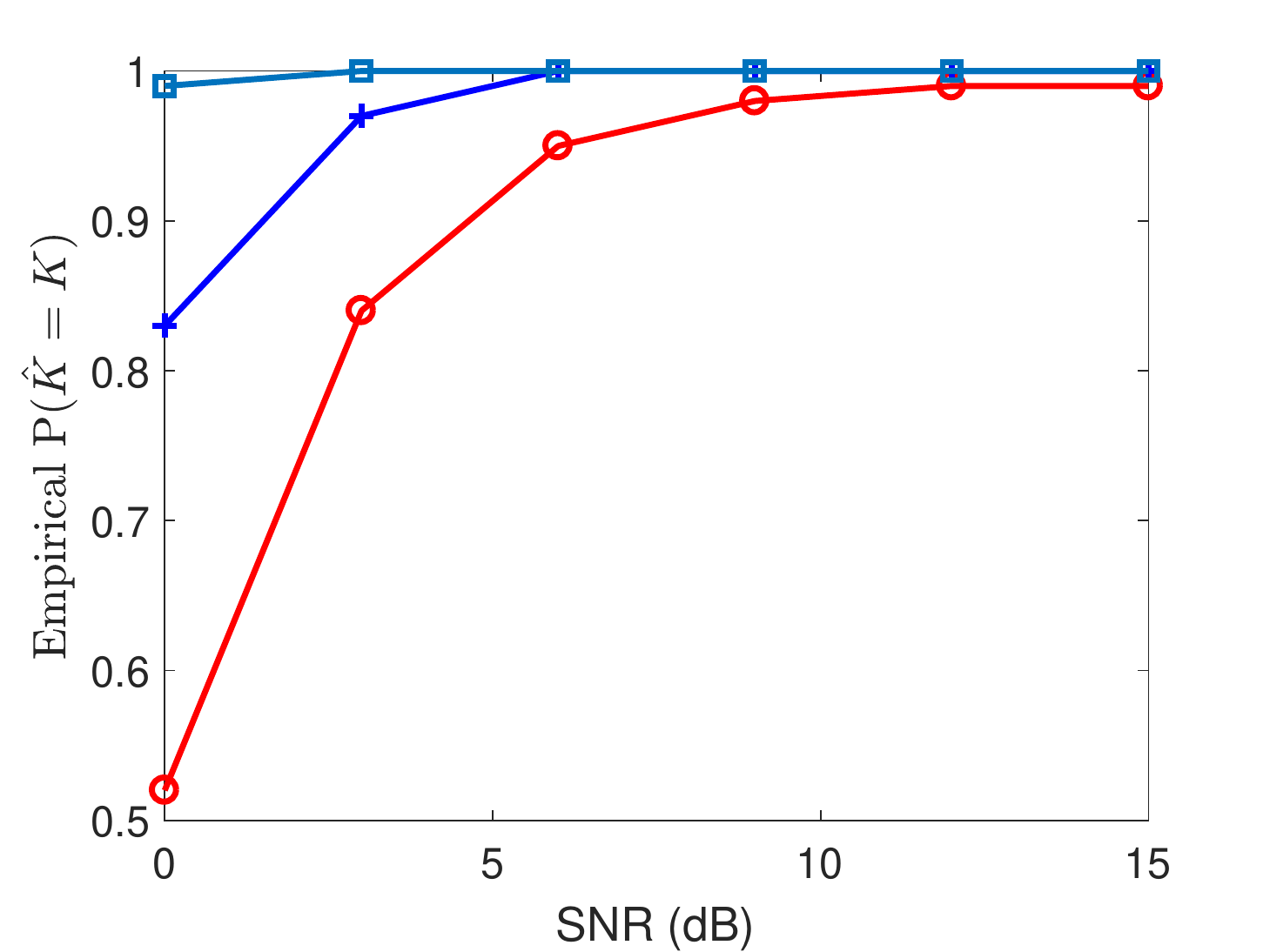}}
  \subfigure[]{
    \label{Case2SNRgen4}%% label for first subfigure
    \includegraphics[width=50mm]{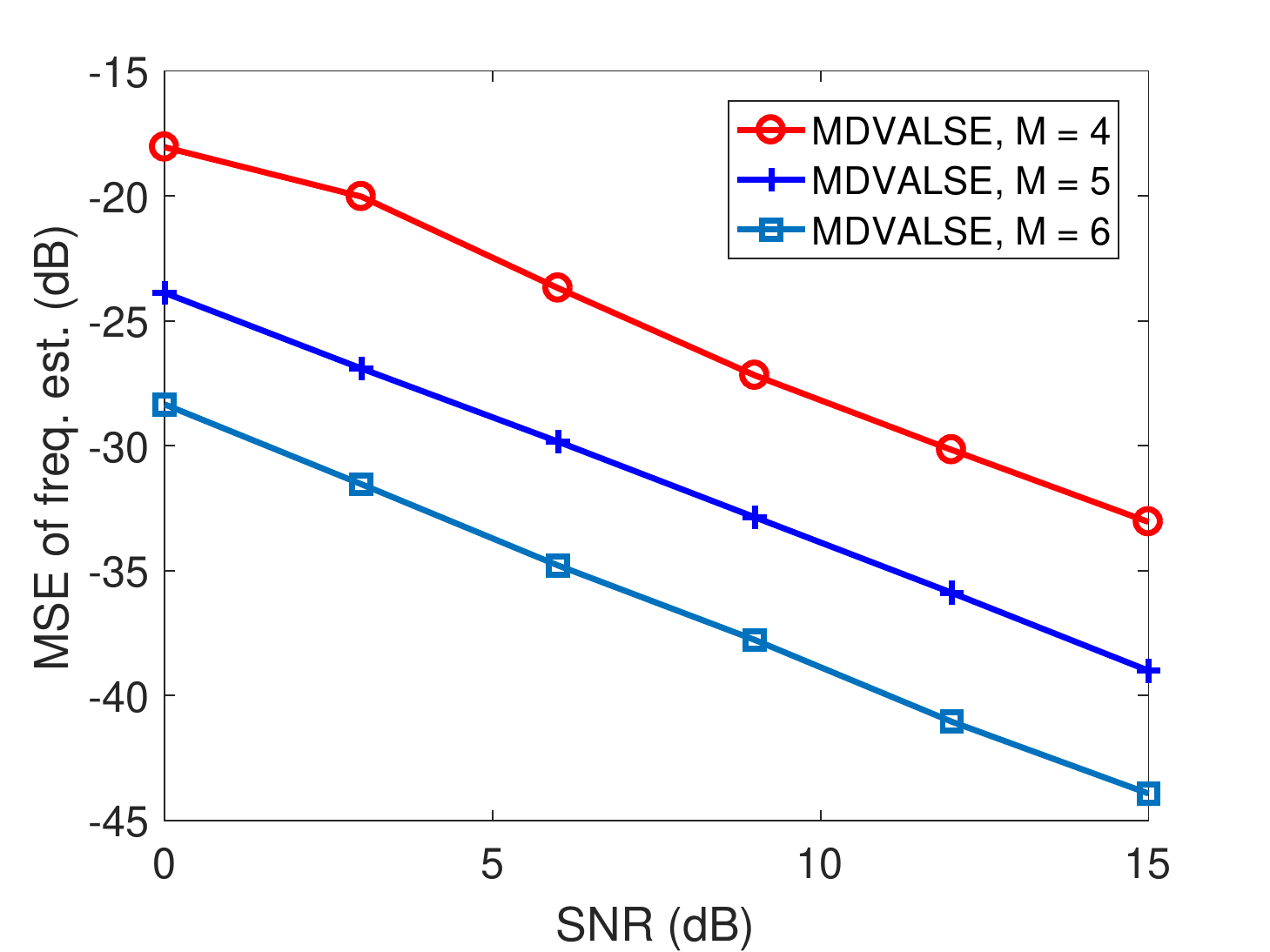}}
 \caption{Performance of MDVALSE for 4D case.}
  \label{4Dcase} %% label for entire figure
\end{figure*}

Now we analyze the computation complexity of MDVALSE. The complexity of initialization is dominated by $D$ dimensional FFT. For the $i$th dimension, the number of grid is $\gamma M_i$ where $\gamma$ can be viewed as the oversampling factor. Thus the initialization has complexity $\mathcal O(N\prod_{d=1}^DM_d\log(\prod_{d=1}^DM_d))$. For each iteration, the complexity is dominated by the estimate of $\mathbf s$ and the projection of $\{q(\boldsymbol\theta_k|\mathbf Y)\}_{k\in\widehat{\mathcal S}}$ as $D$ independent VM. According to \cite{Badiu}, the calculation of $\widehat{\mathbf s}$ has complexity ${\mathcal O}(N{\widehat K}^3)$ and in general $N=\min_{d}M_d$. For the projection operation, the complexity is ${\mathcal O}({\widehat K}D^2\prod^D_{d=1}M_d+{\widehat K}D^3)$ which are mainly dominated by the calculation and inverse of Hessian matrix in (\ref{kappa_post}). Therefore, the computational complexity of MDVALSE is $\mathcal O(N\prod_{d=1}^DM_d\log(\prod_{d=1}^DM_d)+T({\widehat K}D^2\prod^D_{d=1}M_d+{\widehat K}D^3))$ where $T$ denotes the number of whole iterations. For $M_1=M_2\cdots=M_D=M$, the computation complexity can be simplified as $\mathcal O(NDM^{D}\log M + T(N{\widehat K}^3 + {\widehat K}D^2M^D+{\widehat K}D^3))$  .
\section{Numerical Simulation}\label{NS}
In this section, substantial numerical experiments are conducted to evaluate the performance of MDVALSE algorithm. The normalized mean squared error (NMSE) of $\widehat{\mathbf X}$, MSE of $\widehat{\boldsymbol\theta}$ and the correct model order estimated probability ${\rm P}(\widehat{K}=K)$ are adopted as the performance metrics. The frequencies estimation error is averaged over the trials in which ${\widehat K} = K$ for a given simulation point. We define nominal signal-to-noise ratio (SNR) as ${\rm SNR} \triangleq 10{\rm log}(\parallel{\mathbf X}\parallel_{\rm F}^2/\parallel{\mathbf N}\parallel_{\rm F}^2$).
The Algorithm stops when  $||\widehat{\mathbf X}^{(t-1)} - \widehat{\mathbf X}^{(t)}||_{\rm F}/||\widehat{\mathbf X}^{(t-1)}||_{\rm F}< 10^{-6}$ or $t > 500$, where $t$ is the number of iterations.

At first, the approximations of noncohent PDF (\ref{nonconpdf}) and coherent PDF (\ref{pdf-q}) as $D$ independent von Mises distribution are validated numerically and results are shown in Fig. \ref{approx}. Note that for $\gamma=1.6$, the approximation deviates a little away from the true PDF due to the inaccuracy of the FFT results. As $\gamma$ increases to $52$, the approximation is high and the two PDFs are indistinguishable around the peak.
\subsection{Estimation from 2D case}
In this simulation, we consider a 2D case with $K=8$ signals and true frequencies are $[(0.94,1.26),\\(1.26,-2.51),(1.89,1.89),(2.83,-1.26),
(-2.51,1.57),(-2.51,-2.51),(-1.57,2.51),(-1.45,2.76)].$
The complex weight coefficients are generated i.i.d. from complex normal distribution ${\mathcal {CN}}(0,1)$. The number of measurements is $M_1 = M_2 = 10$. As for performance comparison,  reweighted trace minimization (RWTM) algorithm proposed in \cite{yangzaiMD} is chosen and the performances of RWTM and MDVALSE are investigated in Fig. \ref{2Dcase}.
It can be seen that for noiseless case in Fig. \ref{2Dcasea}, MDVALSE and RWTM algorithms both estimate well. Numerically, the MSE of frequencies of RWTM is lower than that of MDVALSE. For noisy case in Fig. \ref{2Dcaseb}, MDVALSE still performs well, while RWTM overestimates the model order. This demonstrates that MDVALSE is more robust against noise than RWTM.
\subsection{Estimation from 4D case}
The performance of MDVALSE algorithm for 4D case versus SNR is investigated. The magnitudes and phases of the complex weight coefficients are generated i.i.d. from normal distribution $\mathcal N(1,0.2)$ and uniform distribution $\mathcal U(-\pi,\pi)$, respectively. The number of sinusoids is $K = 3$ and the wrap-around distance between any two frequencies is at least $\frac{2\pi}{N}$ radians for each dimension. $100$ MC trails are performed and results are shown in Fig. \ref{4Dcase}. MDVALSE works well and performs better as SNR or number of measurements increases.
\section{Conclusion}
This paper develops the MDVALSE for multidimensional line spectral estimation, where the multidimensional frequencies are treated as a whole and their PDF are  projected as independent von Mises distribution for tractable inference. MDVALSE inherits the advantages of automatically estimating the model order, number of nuisance parameters such as noise variance, and providing the uncertain degree of frequency estimates. Numerical results demonstrate the effectiveness of MDVALSE.

\end{document}